\documentclass[prd,twocolumn,showpacs,preprintnumbers,amsmath,amssymb]{revtex4}

\usepackage[active]{srcltx}
\usepackage{amsthm}
\usepackage{latexsym}
\usepackage{mathrsfs}
\usepackage{eufrak}
\usepackage{graphicx}
\usepackage{dcolumn}
\usepackage{bm}

\theoremstyle{plain}

\newtheorem{lemma}{Lemma}
\newtheorem{theorem}{Theorem}

\newtheorem{example}{Example}

\begin{document}

\preprint{APS/123-QED}

\title{Initial data sets for the Schwarzschild spacetime}
\author{Alfonso Garc\'{\i}a-Parrado G\'omez-Lobo}
\email{algar@mai.liu.se}
 \affiliation{Matematiska institutionen, Link\"opings Universitet,
SE-58183 Link\"oping, Sweden.}
\author{Juan A. Valiente Kroon}%
 \email{j.a.valiente-kroon@qmul.ac.uk}
\affiliation{School of Mathematical Sciences, Queen Mary, University of London,
Mile End Road, London E1 4NS, UK.}%

\date{\today}

\begin{abstract}
A characterisation of initial data sets for the Schwarzschild
spacetime is provided. This characterisation is obtained by performing
a 3+1 decomposition of a certain invariant characterisation of the
Schwarzschild spacetime given in terms of concomitants of the Weyl
tensor. This procedure renders a set of
necessary conditions ---which can be written in terms of the electric
and magnetic parts of the Weyl tensor and their concomitants--- for an
initial data set to be a Schwarzschild initial data set. Our approach
also provides a formula for a static Killing initial data set
candidate ---a KID candidate. Sufficient conditions for an initial
data set to be a Schwarzschild initial data set are obtained by
supplementing the necessary conditions with the requirement that the
initial data set possesses a stationary Killing initial data set of the
form given by our KID candidate. Thus, we obtain an algorithmic
procedure of checking whether a given initial data set is
Schwarzschildean or not.
\end{abstract}

\maketitle
 \section{Introduction}
The Schwarzschild and Kerr spacetimes occupy a privileged role among
the asymptotically flat solutions to the Einstein field
equations. Although they are respectively, static and stationary, they
are models for dynamical black hole spacetimes. Moreover, they are
believed to describe, in a certain sense, the asymptotic state of
the dynamical black hole spacetimes.

It is generally agreed that in the next years, the main source of
understanding of dynamical black hole spacetimes will come from
numerical simulations of these solutions to the Einstein field
equations. Normally, these numerical simulations make use of a 3+1
splitting of the spacetime. Thus, it becomes of relevance acquiring an
understanding of the 3+1 features of the Schwarzschild and
---eventually--- the Kerr solutions. 
Some relevant research in this direction has been carried out 
in \cite{BeiOMu98,BriCavIse80,OMuMal03,ParFra06,Rei73,
SchBauCooShaTeu98,Sch02,Val04c,Val05b}. 

When performing a 3+1 decomposition of the Schwarzschild spacetime
---or essentially any other exact solution to the Einstein
equations--- the relative simplicity of the solution gets blurred by
the choice of gauges which do not reflect the symmetries of the
spacetime or its algebraic structure ---as in the case of the  Petrov
type. In this context, invariant characterisations of the spacetime
become very useful ---in particular those involving scalars.

Precisely in this set of mind, in reference \cite{Val05b}, the
following question was raised: when an initial data set for the
Einstein vacuum field equations corresponds to a slice in the
Schwarzschild spacetime?  By \emph{the Schwarzschild spacetime} it
will be understood the Schwarzschild-Kruskal maximal extension,
$(\mathcal{M},g_{\mu\nu})$, of the Schwarzschild spacetime
\cite{Kru60}.  An initial data set for the Einstein vacuum field
equations is a triplet $(\mathcal{S},h_{ij},K_{ij})$ consisting of a
Riemannian manifold $(\mathcal{S},h_{ij})$ and a symmetric tensor
field $K_{ij}$ in $\mathcal{S}$ ---the first and second fundamental
form, or alternatively, the initial 3-metric and extrinsic
curvature--- satisfying the Einstein vacuum constraint equations.
Under these circumstances the manifold $\mathcal{S}$ can be
isometrically embedded in a four dimensional Lorentzian manifold
$(\mathcal{M},g_{\mu\nu})$ ---a spacetime--- which is a solution of
Einstein vacuum field equations.  Such spacetime is called the {\em
  development} of the initial data set.  It is well-known that a {\em
  maximal development} of an initial data set exists and is unique up
to isometry \cite{ChoGer69}.  In this paper we establish conditions
under which the development of an initial data set is a subset of
Schwarzschild spacetime.  Data sets satisfying this property are called
Schwarzschild initial data.  We shall assume $\mathcal{S}$ to be
either a Cauchy hypersurface or an hyperboloid ---that is, a spacelike
hypersurface which is asymptotically null. It is noted that in
reference \cite{OMuRos06}, a similar question, restricted to the case
of spherically symmetric 3-geometries was raised.

A partial answer to the question of the characterisation of
Schwarzschild initial data sets was provided in
\cite{Val05b}. It made use of the fact that for a vacuum spacetime,
$(\mathcal{M},g_{\mu\nu})$, the relation
\begin{equation}
\nabla^\sigma \nabla_\sigma C_{\mu\nu\lambda\rho}=f C_{\mu\nu\lambda\rho}, 
\label{Zakharov}
\end{equation}
which in the sequel we will call the \emph{Zakharov property}, where
$C_{\mu\nu\lambda\rho}$ denotes the Weyl tensor of $g_{\mu\nu}$ and
$f\neq0$ is a function, implies that the spacetime is of Petrov type D
\cite{Zak72}. All the vacuum type D spacetimes are known, so a direct
calculation reveals that the only Petrov type D spacetimes satisfying
the Zakharov property are those which have an hypersurface orthogonal
timelike Killing vector ---thus, for example the Kerr solution does
not satisfy the Zakharov property, but the Schwarzschild spacetime
does. The Zakharov property projects nicely under a 3+1 decomposition,
making it amenable for a characterisation of Schwarzschild initial
data. The characterisation given in \cite{Val05b} was completed by the
addition of two extra ingredients: the asymptotic flatness
---asymptotic Euclideanity--- of the initial data and the
non-vanishing of the ADM mass. An important point which was left open
was the question of the propagation of the Zakharov condition: does
the development of an initial data satisfying the Zakharov property
satisfy the property at latter times?

In the present article we pursue an alternative characterisation of
Schwarzschild data. In particular, we want to avoid the use of the two
global ingredients used in \cite{Val05b}: the asymptotic flatness and
the non-vanishing of the ADM mass. Our starting point is a certain
invariant characterisation of the Schwarzschild spacetime in terms of
concomitants of the Weyl tensor which was given in \cite{FerSae98}.
Necessary conditions for a pair $(h_{ij},K_{ij})$ satisfying the
vacuum constraint equations to be Schwarzschild data are obtained by
making a 3+1 splitting of the characterisation of reference
\cite{FerSae98}. Now, in order to answer the more subtle question
whether the necessary conditions are also sufficient ones, one has to
confront in one way or the other the issue of their {\em propagation}.
In our context the propagation of the conditions resulting from the
3+1 splitting of the characterisation of \cite{FerSae98} would imply
that the initial data under question are indeed Schwarzschildean.  If
the conditions do not propagate, then we may need to add extra
conditions in order to construct Schwarzschild initial data.  This
issue is dealt with by observing that the invariant characterisation
of \cite{FerSae98} also yields a formula for the static Killing vector
of the Schwarzschild spacetime. This formula is again given in terms
of concomitants of the Weyl tensor, and can be split in \emph{lapse}
and \emph{shift} terms. These shift and lapse can be pull-backed to
the manifold $\mathcal{S}$ and if they satisfy certain conditions
which we shall call \emph{the Killing initial data conditions} ---the
KID conditions--- then the development of the initial data possesses a
Killing vector. Now, if the restriction of the resulting Killing
vector to the initial hypersurface $\mathcal{S}$ is timelike for an open 
subset $\mathcal{U}\subset\mathcal{S}$, then by continuity, there will
be a timelike Killing vector in, at least, a small portion of the
domain of dependence of $\mathcal{U}$, $\mathcal{D}(\mathcal{U})$.

We can take advantage of the existence of a static timelike Killing
vector in $\mathcal{D}(\mathcal{U})$ to show that the conditions
obtained from the 3+1 splitting of the invariant characterisation
mentioned above can be propagated, at least in the small portion of
$\mathcal{D}(\mathcal{U})$, where the Killing vector is stationary,
thereby obtaining a full set of conditions ensuring that this part of
$\mathcal{D}(\mathcal{U})$ is a subset of Schwarzschild
spacetime. This full set of conditions consists of the 3+1 splitting
of the Schwarzschild invariant characterisation and the KID
conditions. A novelty in our approach lies in providing an explicit
Ansatz ---\emph{a Killing candidate}--- for the the solution of these
KID conditions, thus yielding an algorithmic procedure to verify the
existence of a timelike vector in the development of the initial data.

\bigskip
The article is structured as follows: in section \ref{local-char} we
detail the local invariant characterisation of the Schwarzschild spacetime
presented in \cite{FerSae98}. This will be our starting point. Section
\ref{1+3} deals with essential concepts used in the formulation of the
initial value problem in General Relativity and the 3+1 splitting of
geometric quantities. Using these tools we explain in section
\ref{decompositions} how to decompose the characterisation of Ferrando
and S\'aez which leads us to a set of necessary conditions for
initial data to be a Schwarzschild initial data set ---see theorem 
\ref{first_necessary}. A further analysis of these conditions is
performed in section \ref{analysis} where
an algebraic classification of the electric and the
magnetic parts of Weyl tensor
is performed providing canonical forms for them in each
type of Schwarzschild initial 
data set. In section \ref{sufficient} we investigate how to
enlarge the set of necessary conditions in order to obtain a set of
geometric conditions guaranteeing that the initial data are indeed 
Schwarzschild initial data. These extra conditions are given by the
KID equations. The full set of conditions are put together in theorem
\ref{main_result} which is the most important result of this paper. A
practical application of theorem \ref{main_result} is presented in
appendix \ref{kid_ts}.  

A great deal of the algebraic calculations of this paper were undertaken
with the 
MA\-THE\-MATICA package {\em xTensor} \cite{xtensor05}. This package
is specially adapted to calculations in\-vol\-ving abstract 
indexes and among its many
features it includes algorithms to handle efficiently the 3+1 decomposition.

\section{A local characterisation of the Schwarzschild spacetime}
\label{local-char}
We begin by introducing some notation and by fixing
conventions. As in the introduction, $(\mathcal{M},g_{\mu\nu})$
denotes a 4-dimensional spacetime. We shall use the signature
$(-+++)$. The Greek indices $\mu,\; \nu,\ldots$ are abstract spacetime indices
while the Latin indices $i,\;j,\ldots$ are abstract spatial ones.
Boldface Latin indices $\boldsymbol a,\;\boldsymbol b,\ldots$ will be
reserved to denote components with respect to a frame and they will range 
$0,\dots,3$ in a spacetime frame and $1,2,3$ in a spatial frame.

Let $U_{\mu\nu\lambda\psi}$ and $V_{\mu\nu\lambda\psi}$ be two rank 4
tensors antisymmetric in the first and second pair of indices and
symmetric under the interchange of these pair of indices. We define the
$\star$-product of them via \footnote{These definitions of stem from the theory of 2-forms. 
The natural metric in the space of forms is given by the rank 4 tensor 
$\frac{1}{8}(g\wedge g)_{\mu\nu\lambda\psi}$. Thus, in terms of 
this tensor, one has that
\[ (U\star V)_{\mu\nu\lambda\psi}=\frac{1}{8}(g\wedge g)_{\kappa\pi\sigma\tau} 
U_{\mu\nu}^{\ \ \kappa\pi}V_{\lambda\psi}^{\ \ \sigma\tau}
=\frac{1}{2}U_{\mu\nu\kappa\pi}V^{\kappa\pi}_{\ \ \ \lambda\psi}\]
Similarly, one takes traces according to
\[ \mbox{tr\;}U=\frac{1}{8}(g \wedge g)_{\mu\nu\lambda\psi}U^{\mu\nu\lambda\psi}
=\frac{1}{2}U^{\mu\nu}_{\ \ \ \mu\nu}.\]
},
\begin{equation}
(U\star V)_{\mu\nu\lambda\psi}=\frac{1}{2} U_{\mu\nu}^{\phantom{\mu\nu}\kappa\pi}V_{\kappa\pi\lambda\psi}.
\end{equation}

Let $C_{\mu\nu\lambda\psi}$ denote the Weyl tensor associated to the metric $g_{\mu\nu}$. Then
\begin{subequations}
\begin{eqnarray}
&&(C\star C\star C)_{\mu\nu\lambda\psi}=\frac{1}{4}C_{\mu\nu}^{\phantom{\mu\nu}\sigma\tau}C_{\sigma\tau}^{\phantom{\sigma\tau}\kappa\pi}C_{\kappa\pi\lambda\psi}, \\
&&\mbox{tr}(C\star C \star C)=\frac{1}{2}(C\star C \star C)^{\mu\nu}_{\phantom{\mu\nu}\mu\nu}.
\end{eqnarray}
\end{subequations}
Crucial objects in the subsequent analysis are the following scalars:
\begin{subequations}
\begin{eqnarray}
&&\rho=\left( \frac{1}{12}\mbox{tr}(C \star C \star C) \right)^{1/3}, \label{rho}\\
&& \alpha=\frac{1}{9\rho^2}g^{\mu\nu}\nabla_\mu \rho \nabla_\nu\rho+2\rho. \label{alpha} 
\end{eqnarray}
\end{subequations}
Finally, we also write
\begin{equation}
(g\wedge g)_{\mu\nu\lambda\sigma}=2(g_{\mu\lambda}g_{\nu\sigma}-g_{\mu\sigma}g_{\nu\lambda}),
\end{equation}
and define the following tensors:
\begin{subequations}
\begin{eqnarray}
&& S_{\mu\nu\lambda\sigma}=\frac{1}{3\rho}\left(C_{\mu\nu\lambda\sigma}+\frac{1}{2}\rho(g\wedge g)_{\mu\nu\sigma\lambda} \right), \\
&& P_{\mu\nu}=C^*_{\lambda\mu\sigma\nu}\nabla^\lambda\rho\nabla^\sigma\rho \\
&& Q_{\mu\nu}=S_{\lambda\mu\sigma\nu}\nabla^\lambda\rho\nabla^\sigma\rho.
\end{eqnarray}
\end{subequations}
Here, and in the sequel, $C^*_{\mu\nu\lambda\sigma}$ denotes the \emph{(Hodge) dual} of
the Weyl tensor,
\begin{equation}
C^*_{\mu\nu\lambda\sigma}=\frac{1}{2}C_{\mu\nu\kappa\pi}\epsilon^{\kappa\pi}_{\phantom{\kappa\pi}\lambda\sigma}.
\end{equation}

In terms of the above concomitants of the Weyl tensor, Ferrando \&
S\'aez's characterisation of the Schwarzschild spacetime states the
following ---see \cite{FerSae98}---. Note that there are slight differences
between our conventions in the definition of the concomitants and 
those of Ferrando \& S\'aez.
\begin{theorem}[Ferrando \& S\'aez, 1998] \label{thm_FerSae}
Necessary and sufficient conditions for a vacuum spacetime
$(\mathcal{M},g_{\mu\nu})$ to be locally isometric to the
Schwarzschild spacetime are
\begin{subequations}
\begin{eqnarray}
&&\rho\neq 0, \label{type_Da}\\
&& (S\star S)_{\mu\nu\sigma\lambda}-S_{\mu\nu\sigma\lambda}=0, 
\label{type_Db}\\
&& P_{\mu\nu}=0, \label{type_Dc}\\
&& 2 Q_{\mu\nu}v^\mu v^\nu + Q^\mu_{\phantom{\mu}\mu}\leq 0,\label{type_Dd} \\
&& \alpha >0, \label{type_De}
\end{eqnarray}
\end{subequations}
where $v^\mu$ is an arbitrary unit timelike vector. Moreover, the Schwarzschild mass is given by
\begin{equation}
m=\frac{\rho}{\alpha^{3/2}}, \label{mass}
\end{equation}
and a timelike Killing vector in one of the static regions by
\begin{equation}
\xi^\mu=\frac{1}{\rho^{4/3}\sqrt{|Q_{\lambda\nu}v^\lambda v^\nu}|}Q^\mu_{\phantom{\mu}\nu}v^\nu \label{killing}.
\end{equation}
\end{theorem}

\bigskip
\textbf{Remarks.} 
\begin{enumerate}
\item The conditions (\ref{type_Da}) and (\ref{type_Db})
are equivalent (locally) to having static, type D spacetimes ---these were subdivided by Ehlers 
\& Kundt \cite{EhlKun62} into the classes A, B and C; the class A containing the Schwarzschild metric. If in
addition one has that $P_{\mu\nu}\neq 0$, then the spacetime
corresponds to the C-metric ---the only representative of the class C. 
On the other hand, if 
$P_{\mu\nu}=0$ and $ 2
Q_{\mu\nu}v^\mu v^\nu + Q^\mu_{\phantom{\mu}\mu}>0$ one has a type A
metric. The condition $\alpha>0$ is the one that selects the Schwarzschild 
metric among 
the 3 contained in the Ehlers-Kundt class A.

\item  Theorem \ref{thm_FerSae} can be reformulated  replacing condition
(\ref{type_Dd}) by 
\begin{equation}
Q_{\mu\nu}=-9\alpha\rho^2\xi_\mu\xi_\nu \tag{\ref{type_Dd}'}.
\end{equation} 
where $\xi_\mu$ is a smooth 
vector field which is a Killing vector in those regions
where it is timelike.
To show this we first note that (\ref{type_Dd}') entails (\ref{type_Dd})
and hence $\mathcal M$ is locally Schwarzschild. Reciprocally if 
$\mathcal M$ is a subset of Schwarzschild spacetime then,
an explicit computation in a coordinate system covering 
$\mathcal M$ enables us to derive (\ref{type_Dd}'). In our 
calculation we chose the coordinate chart given in 
 \cite{Klosch} which covers the full 
Schwarzschild-Kruskal analytical extension and hence it can be used 
to construct a coordinate patch covering any subset $\mathcal M$ of such
extension. In this paper we shall use theorem \ref{thm_FerSae} with
condition (\ref{type_Dd}') instead of condition (\ref{type_Dd}) because
the former is better suited for our calculations.

\end{enumerate}

\section{Standard results from the $3+1$ decomposition}
\label{1+3}
As mentioned in the introduction, the first part of our analysis will
be concentrated with obtaining a 3+1 splitting of the conditions given in
theorem \ref{thm_FerSae}. Let $(\mathcal{S},h_{ij})$ be a 3-dimensional
connected Riemannian manifold.  
The map $\phi:\mathcal{S}\longrightarrow\mathcal{M}$ is an isometric
embedding if $\phi^*g_{\mu\nu}=h_{ij}$ where as usual
$\phi^*$ denotes the
pullback of tensor fields from $\mathcal{M}$ to $\mathcal{S}$.
In the framework of the 3+1 decomposition it is more advantageous 
to work with {\em foliations}. This is a family of submanifolds
$\{S_t\}_{t\in I}$ of $\mathcal{M}$ whose union is an open 
subset ${\mathcal N}\subset\mathcal{M}$ and no two 
submanifolds of the family have common points. Each submanifold of the 
foliation is called a leaf. When ${\mathcal N}={\mathcal M}$ then 
$\{S_t\}_{t\in I}$ is called a foliation of $\mathcal M$.
The foliations relevant for us are those formed by 
leaves which are 3-dimensional spacelike hypersurfaces of $\mathcal M$. 
In this case we 
can define a unit 1-form $n_\mu$ on $\mathcal{N}$ by the property 
that $n_\mu$ is normal to any of the leaves of the foliation.
Since the leaves are spacelike we have $n^\mu n_\mu=-1$. 
The tensor fields $h_{\mu\nu}$ and $K_{\mu\nu}$ of $\mathcal{N}$  
defined by the relations
\begin{equation}
h_{\mu\nu}\equiv g_{\mu\nu}+n_{\mu}n_{\nu},\ 
K_{\mu\nu}\equiv-\frac{1}{2}\mathcal{L}_{n}h_{\mu\nu},
\label{lie-h}
\end{equation}
play the role of the first and second fundamental form 
respectively for any of the leaves. Here 
$\mathcal{L}$ denotes the Lie derivative.
From now on we will assume that 
$\phi(\mathcal{S})$ belongs to a given foliation. 
It is then clear that $\phi^*h_{\mu\nu}=h_{ij}$. 
 Further, we may define the tensor field
$K_{ij}\equiv\phi^*K_{\mu\nu}$. 

The $3+1$ splitting of a timelike vector field $t^\mu$ 
can be also easily achieved in terms of its  
\emph{lapse}, $N$, and \emph{shift}, $N^\mu$:
\begin{equation}
t^\mu=N n^\mu +N^\mu,\ N\equiv-t_{\mu}n^\mu,\ N^\mu\equiv h^\mu_{\ \nu}t^\nu.
\end{equation}

Less trivially, we consider now the $3+1$ splitting of the Weyl tensor 
---and its dual--- in terms of its electric and magnetic parts 
with respect to $n^\mu$:
\begin{subequations}
\begin{eqnarray}
& &C_{\mu\nu\lambda\sigma}=2\left(l_{\mu[\lambda}E_{\sigma]\nu}-l_{\nu[\lambda}E_{\sigma]\mu}-n_{[\lambda}B_{\sigma]\tau}\epsilon^\tau_{\phantom{\tau}\mu\nu}
-\right.\nonumber\\
& &\left.-n_{[\mu}B_{\nu]\tau}\epsilon^\tau_{\phantom{\tau}\lambda\sigma} 
\right),\label{weyl-1} \\
& &C^*_{\mu\nu\lambda\sigma}=2\left(l_{\mu[\lambda}B_{\sigma]\nu}-l_{\nu[\lambda}B_{\sigma]\mu}+n_{[\lambda}E_{\sigma]\tau}\epsilon^\tau_{\phantom{\tau}\mu\nu}
+\right.\nonumber\\
& &\left.+n_{[\mu}E_{\nu]\tau}\epsilon^\tau_{\phantom{\tau}\lambda\sigma} \right),\label{weyl-2}
\end{eqnarray}
\end{subequations}
where
\begin{equation}
E_{\tau\sigma}\equiv C_{\tau\nu\sigma\lambda}n^\nu n^\lambda, \quad 
B_{\tau\sigma}\equiv C^*_{\tau\nu\sigma\lambda}n^\nu n^\lambda,
\label{e-m}
\end{equation}
denote the \emph{n-electric} and \emph{n-magnetic} parts respectively,
and $l_{\mu\nu}\equiv h_{\mu\nu}+n_{\mu}n_{\nu}$, while
$\epsilon_{\tau\lambda\sigma}\equiv\epsilon_{\nu\tau\lambda\sigma}n^{\nu}$. 
The tensors $E_{\mu\nu}$ and
$B_{\mu\nu}$ are symmetric, traceless, and spatial: $n^\mu
E_{\mu\nu}=n^\mu B_{\mu\nu}=0$. 
The tensor $\epsilon_{\mu\nu\sigma}$ is
fully antisymmetric and when a foliation is present it can 
be put in correspondence with
the volume element of any of its leaves.

Let $D_\mu$ denote the operator
obtained from the spacetime connection $\nabla_\mu$ by:
\begin{equation}
D_\mu T^{\alpha_1\dots\alpha_p}_{\ \beta_1\dots\beta_q}\equiv
h^{\alpha_1}_{\ \rho_1}\dots h^{\alpha_p}_{\ \rho_p}h^{\sigma_1}
_{\ \beta_1}\dots
h^{\sigma_q}_{\ \beta_q}h^{\lambda}_{\ \mu}\nabla_\lambda
T^{\rho_1\dots\rho_p}_{\ \sigma_1\dots\sigma_q},
\end{equation}
$p,q\in\mathbb{N}$, where $T^{\alpha_1\dots\alpha_p}_{\
  \beta_1\dots\beta_q}$ is any tensor field in $\cal M$.  We have the
important property:
\begin{equation}
\phi^*(D_\mu T^{\alpha_1\dots\alpha_p}_{\ \beta_1\dots\beta_q})=
D_i(\phi^*T^{\alpha_1\dots\alpha_p}_{\ \beta_1\dots\beta_q}),
\end{equation}
the operator $D_i$ being the Levi-Civita connection
associated with the 3-metric $h_{ij}$:
\begin{equation}
D_jh_{ik}=0.
\end{equation}
For later use we need to find the $3+1$ splitting of $\nabla_\mu
E_{\mu\lambda}$ and $\nabla_\mu B_{\mu\lambda}$ when Einstein vacuum
equations hold. This can be done by finding the 3+1 splitting of
the contracted Bianchi identity $\nabla_\rho C^\rho_{\ \mu\nu\sigma}=0$,
valid when $R_{\mu\nu}=0$ and use (\ref{weyl-1}), (\ref{weyl-2}). As is 
well-known (see e.g. \cite{Fri96}) the  
result is ($a^\mu\equiv n^\rho\nabla_\rho n^\mu$) 
\begin{widetext}
\begin{subequations}
\begin{eqnarray}
&&\mathcal{L}_nE_{\mu\nu}^{{  }}
=2a_\lambda B_{{\sigma(\nu}}^{{  }} 
\varepsilon_{{\mu)  }}^{{\ \sigma\lambda}}-3 E_{{\nu }}^{{\ \sigma}} 
K_{{\mu\sigma}}^{{  }}-2 E_{{\mu }}^{{\ \sigma}} K_{{\nu\sigma}}^{{  }}
+2 E_{{\mu\nu}}^{{  }} K_{{\ \sigma}}^{{\sigma }}
+E_{{  }}^{{\sigma\lambda}} K_{{\sigma\lambda}}^{{ }} h_{{\mu\nu}}^{{  }}
+\varepsilon_{{\mu  }}^{{\ \sigma\lambda}}(D_\lambda B_{{\nu\sigma}}^{})\label{lie-e}\\
&&\mathcal{L}_nB_{{\mu\nu}}^{{  }}=-2a_\lambda E_{{\sigma(\nu}}^{{  }} 
\varepsilon_{{\mu)  }}^{{\ \sigma\lambda}}-3 B_{{\nu }}^{{\ \sigma}} 
K_{{\mu\sigma}}^{{  }}-2 B_{{\mu }}^{{\ \sigma}} K_{{\nu\sigma}}^{{  }}
+2 B_{{\mu\nu}}^{{  }} K_{{\ \sigma}}^{{\sigma }}+B_{{  }}^{{\sigma\lambda}} 
K_{{\sigma\lambda}}^{{ }} h_{{\mu\nu}}^{{  }}
-\varepsilon_{{\mu  }}^{{\ \sigma\lambda}} (D_\lambda E_{{\nu\sigma}}^{{  }}),
\label{lie-b}
\end{eqnarray}
\end{subequations}
\end{widetext}
which are {\em evolution equations} and 
\begin{subequations}
\begin{eqnarray}
&& D^j E_{ji}+B_{jk} K^j_{\phantom{j}l}\epsilon^{kl}_{\phantom{kl}i}=0, \\
&& D^j B_{ji}-E_{jk} K^j_{\phantom{j}l}\epsilon^{kl}_{\phantom{kl}i}=0,
\end{eqnarray}
\end{subequations}
which are constraint equations. Expanding the Lie derivative in terms of 
the Levi-Civita covariant derivative we find the desired 3+1 decomposition 
of $\nabla_\mu E_{\mu\lambda}$ and $\nabla_\mu B_{\mu\lambda}$.
\subsubsection{Initial data sets for the Vacuum equations}
As explained in the
introduction, $(\mathcal{S},h_{ij},K_{ij})$ is an initial data set for
the vacuum Einstein field equations if $h_{ij}$, $K_{ij}$ satisfy the
constraint equations:
\begin{subequations}
\begin{eqnarray}
&& r+ K^2-K^{ij}K_{ij}=0, \label{Hamiltonian}\\
&& D^jK_{ij}-D_iK=0, \label{Momentum}
\end{eqnarray}
\end{subequations} 
on $\mathcal{S}$ where $r$ is the Ricci scalar of $h_{ij}$, and we
write $K=K^i_{\phantom{i}i}$. If $(\mathcal{M},g_{\mu\nu})$ is the
initial data development and $\phi:\mathcal{S}\rightarrow\mathcal{M}$
the isometric embedding of the initial data, then we have the property 
$K_{ij}=\phi^*K_{\mu\nu}$,
where $K_{\mu\nu}$ is defined by (\ref{lie-h})
(of course a foliation of $\mathcal M$ containing $\phi({\mathcal S})$ 
must be constructed first).
From the initial data $(\mathcal{S},h_{ij},K_{ij})$ we may define 
the tensor fields $E_{ij}$, $B_{ij}$ by the relations
\begin{subequations}
\begin{eqnarray}
&& E_{ij}\equiv r_{ij}+K K_{ij}-K_{ik}K^k_{\phantom{k}j}
\label{electric_idata},
\\ && B_{ij}\equiv\epsilon_{(i}^{\phantom{i}kl}D_{|k}K_{l|j)}\
\label{magnetic_idata}.
\end{eqnarray}
\end{subequations}
We note that the symmetrization in the last equation is not needed if
the constraints (\ref{Hamiltonian}) and (\ref{Momentum}) are satisfied
---which we shall assume. Thus, if the vacuum constraints hold then
$E_{ij}$, $B_{ij}$ are both symmetric and traceless. Defining in the
data development the electric and magnetic parts of the Weyl tensor as
shown in (\ref{e-m}) we have $E_{ij}=\phi^*E_{\mu\nu}$, and $\
B_{ij}=\phi^*B_{\mu\nu}$.

\section{3+1 decomposition of the characterisation by Ferrando \& S\'aez}
\label{decompositions}
We proceed now to a 3+1 decomposition of the conditions appearing in the
invariant characterisation of theorem \ref{thm_FerSae}. 
This leads naturally to a first set of necessary conditions for a given initial
data set $(\mathcal{S},h_{ij},K_{ij})$ to be a Schwarzschild initial
data set. We study each condition of the theorem separately in the forthcoming
subsections.
\subsection{Decomposition of equation (\ref{type_Da})}
Firstly, we note that:
\begin{equation}
\mbox{tr}(C\star C \star C)=6 B_{\mu}^{\phantom{\mu}\nu}B^{\mu\sigma}
E_{\nu\sigma}-2 E_{\mu\nu}E_{\sigma}^{\phantom{\nu}\nu}
E^{\mu\sigma}, \label{trCCC}
\end{equation}
from where the scalar $\rho$ ---see equation (\ref{rho})--- can be calculated
yielding
\begin{equation}
\rho=\left(\frac{1}{2} B_{\mu}^{\phantom{\mu}\nu}B^{\mu\sigma}E_{\nu\sigma}
-\frac{1}{6} E_{\mu\nu}E_{\sigma}^{\phantom{\nu}\nu}E^{\mu\sigma} \right)^{1/3}.
\end{equation} 
Using other conditions of theorem \ref{thm_FerSae} we will be 
able to get a simpler expression for $\rho$ ---see (\ref{rhoscalar}) below.

\subsection{Decomposition of equation (\ref{type_Db})}
Using the formula for the Weyl tensor in terms of its electric and
magnetic parts, equation (\ref{weyl-1}), it is not difficult to decompose
(\ref{type_Db}). This decomposition renders the condition
(\ref{type_Db}) of theorem \ref{thm_FerSae} in the form
\begin{widetext}
\begin{subequations}
\begin{eqnarray} 
& &4 {E}_{\mu [\delta }^{{  }} {E}_{\lambda]\nu}^{{  }}+
2 h_{\nu [\lambda }^{{  }} (B_{{\delta] \pi}}^{{  }}B_{{\mu\  }}^{{\ \pi}}+{E}_{{\delta]
\pi}}^{{  }}{E}_{{\mu  }}^{{\ \pi}}-{\rho}_{}{E}_{\delta]\mu}^{{  }})
\nonumber\\
& &\phantom{XXX}+2 h_{\mu[\delta }^{{  }} (B_{{\lambda] }}^{{\ \pi}} B_{{\nu \pi}}^{{  }}
+{E}_{{\lambda]  }}^{{\ \pi}} {E}_{{\nu \pi}}^{{  }}-{\rho}_{}{E}_{\lambda]\nu}^{{  }})+2h_{\mu [\lambda}^{{  }} h_{\delta]\nu}^{{  }} (B_{{\pi\kappa}}^{{  }} B_{{  }}^{{\pi\kappa}}+2 {\rho}_{}^2)=0,\label{easy-1}\\
& &B_{{\nu  }}^{{\ \sigma}} {E}_{{\delta  }}^{{\ \alpha}} \epsilon _{{\lambda \sigma\alpha}}^{{   }}
+B_{{  }}^{{\sigma\alpha}}{E}_{{\nu \sigma}}^{{  }} \epsilon _{{\lambda \delta \alpha}}^{{   }}
-B_{{\nu  }}^{{\ \sigma}} {E}_{{\lambda }}^{{\ \alpha}} \epsilon _{{\delta \sigma\alpha}}^{{   }}
+{\rho}_{}B_{{\nu  }}^{{\ \sigma}} \epsilon _{{\lambda \delta \sigma}}^{{   }}=0,\label{easy-2}\\
& &{E}_{{\lambda  }}^{{\ \alpha}} {E}_{{\mu \alpha}}^{{ }}-B_{{\lambda  }}^{{\ \alpha}} B_{{\mu \alpha}}^{{  }}+{\rho}_{}({E}_{\mu \lambda }^{{  }}-2{\rho}_{} 
h_{\mu \lambda }^{{  }})=0.\label{easy-3}
\end{eqnarray}
\end{subequations}
\end{widetext}
From here we obtain the important relations
\begin{subequations}
\begin{eqnarray}
&&B_{{\mu\nu }}^{{  }}= -\frac{1}{{\rho}_{}}
(B_{{\nu  }}^{{\ \lambda}} {E}_{{\mu\lambda}}^{{  }}+B_{{\mu }}^{{\ \lambda}}
{E}_{{\nu \lambda}}^{{  }}),\label{rule1}\\
&&{E}_{\mu \nu }^{{  }}= \frac{1}{{\rho}_{}}(B_{{\nu  }}^{{\ \lambda}} B_{{\mu \lambda}}^{{  }}
-{E}_{{\nu }}^{{\ \lambda}} {E}_{{\mu \lambda}}^{{  }})
+2{\rho}_{} h_{\mu \nu }^{{  }}.\label{rule2}
\end{eqnarray}
\end{subequations}
The conditions (\ref{rule1}) and (\ref{rule2}) are essentially the
same ones which were obtained in \cite{Val05b} from a 3+1 splitting of
the Zakharov property (\ref{Zakharov}).
  
Taking the trace of condition (\ref{rule1}) on finds that
$E_{\alpha\beta}B^{\alpha\beta}=0$, a well know property of the
Schwarzschild spacetime. The condition (\ref{rule2}) enables us to
obtain a simple expression for $\rho$:
\begin{equation}
{\rho}_{}^2=\frac{({E}_{{\lambda \alpha}}^{{ }} {E}_{{  }}^{{\lambda \alpha}}-B_{{\lambda \alpha}}^{{  }} B_{{  }}^{{\lambda \alpha}}
)}{6},\quad
E_{\lambda\alpha}E^{\lambda\alpha}\geq B_{\lambda\alpha}B^{\lambda\alpha}.
\label{rhoscalar}
\end{equation}
It should be
mentioned that ${E}_{{\lambda \alpha}}^{{ }} {E}_{{ }}^{{\lambda
\alpha}}-B_{{\lambda \alpha}}^{{ }} B_{{ }}^{{\lambda \alpha}}$ and
$E_{\lambda\alpha}B^{\lambda\alpha}$ 
are, respectively, the real and imaginary parts
of $I$, one of the invariants of the Weyl tensor ---see
e.g. \cite{McIAriWadHoe94,ZakMcI97}.
From here, the 4-dimensional covariant derivative of $\rho$ can be obtained
which is
\begin{equation}
\nabla_\mu\rho=\frac{1}{6\rho}(E^{\lambda\alpha}\nabla_\mu E_{\lambda\alpha}-
B^{\lambda\alpha}\nabla_\mu B_{\lambda\alpha}).
\end{equation}
Replacing the derivatives of the electric and magnetic parts by their
$3+1$ splitting (eqs. (\ref{lie-e})-(\ref{lie-b}) must be 
used to find such splitting) we obtain the $3+1$ splitting of
$\nabla_\mu\rho$.  We shall write it in shift and lapse parts as
follows,
\begin{equation}
\nabla_\mu\rho=Pn_\mu +P_\mu,
\label{def-rho}
\end{equation}
where 
\begin{subequations}
\begin{eqnarray}
& &\hspace{-.8cm}P\equiv -\frac{1}{2} {E}_{{  }}^{{\alpha\lambda }}
{K}_{{\alpha\lambda }}^{{  }}
-{\rho}_{} {K}_{{\ \alpha}}^{{\alpha }}\nonumber\\
& &-\frac{1}{6{\rho}_{}}\epsilon _{{\ \ \ \alpha}}^{{\beta\sigma }}
\left({E}_{{  }}^{{\alpha\lambda }} D_\sigma B_{{\lambda \beta}}^{{  }}
+B_{{  }}^{{\alpha\lambda }} D_\sigma{E}_{{\lambda \beta}}^{{  }}\right), \label{scalarP}\\
& &\hspace{-.8cm}P_\mu\equiv\frac{1}{6 {\rho}_{}}(-B_{{  }}^{{\kappa\lambda }}
D_\mu B_{{\kappa\lambda }}^{{  }}
+{E}_{{  }}^{{\kappa\lambda }} D_\mu{E}_{{\kappa\lambda}}^{{  }})=D_\mu\rho.
\label{vectorP}
\end{eqnarray}
\end{subequations}
Equations (\ref{rule1}), (\ref{rule2}) are equivalent to
(\ref{easy-2}) and (\ref{easy-3}). If $B_{\mu\nu}\neq 0$ ---the case
$B_{\mu\nu}=0$ is studied separately in section \ref{pure-e}--- then
(\ref{easy-1}) can be transformed into
\begin{eqnarray}
 & &{E}_{\lambda [\nu }^{{  }} {E}_{\mu] \rho }^{{  }}+ {E}_{{\rho \sigma}}^{{  }}
{E}_{{[\mu  }}^{{\ \sigma}} h_{\nu]\lambda}^{{  }}+
 {E}_{{\lambda  }}^{{\ \sigma}} {E}_{{\sigma [\nu}}^{{  }} h_{\mu] \rho }^{{  }}
\nonumber\\
& &\hspace{5mm}+\frac{1}{2}{E}_{{\sigma\alpha}}^{{  }} {E}_{{  }}^{{\sigma\alpha}} h_{\lambda [\mu }^{{  }} h_{\nu] \rho }^{{  }}=0. \label{long}
\end{eqnarray}
\begin{lemma}
The equation (\ref{long}) is an identity.
\label{lemma-long}
\end{lemma}
\noindent
{\it Proof :} \hspace{3mm}
We start with the identity
\begin{equation}
C_{[\mu\nu}^{\ \ [\alpha\beta}\delta^{\sigma]}_{\ \pi]}=0,
\end{equation}
which holds only in dimension four ---see e.g. \cite{EdHog02}. 
Multiplying this equation by $C^{\mu\nu}_{\ \ \sigma\lambda}$
and expanding we get
\begin{eqnarray*}
& &2 {C}_{{\ \ \ \pi }}^{{\alpha\mu\ \nu}}
   {C}_{{\lambda\mu\ \nu}}^{{\ \ \ \beta}}
+{C}_{{    }}^{{\alpha\beta\mu\nu}}
   {C}_{{\lambda\pi\mu\nu}}^{{    }}\nonumber\\
& &-2 {C}_{{\ \ \ \lambda }}^{{\alpha\mu\ \nu}}
   {C}_{{\ \nu\pi\mu}}^{{\beta   }}+\delta
   _{{\lambda }}^{{\ \beta}} 
{C}_{{      }}^{{\alpha\mu\nu\sigma}}
   {C}_{{\pi\mu\nu\sigma}}^{{    }}
-\delta_{{\ \lambda}}^{{\alpha }} 
{C}_{{      }}^{{\beta\mu\nu\sigma}}
   {C}_{{\pi\mu\nu\sigma}}^{{    }}=0.
\end{eqnarray*}
Replacing the Weyl tensor by (\ref{weyl-1}) we obtain an expression which can 
be written in the form
\begin{equation}
c^{\alpha\beta}_{\ \lambda\pi}+n^{[\alpha} d^{\beta]}_{\ \lambda\pi}+
n_{[\lambda}t^{\alpha\beta}_{\ \pi]}=0,
\end{equation}
where $c^{\alpha\beta}_{\ \lambda\pi}$ is the tensor appearing on the 
right hand side of (\ref{long}) with two indexes raised and 
$d^\beta_{\ \lambda\pi}$, $t^{\alpha\beta}_{\ \pi}$ are spatial tensors. 
The previous equation implies that
 $c^{\alpha\beta}_{\ \lambda\pi}=0$ thus proving the lemma. $\blacksquare$

\subsection{Decomposition of equation (\ref{type_Dc})}
In order to alleviate the notation in our equations it is convenient 
to introduce the following definitions
\begin{equation}
\tilde{E}_\mu\equiv E_{\mu\nu}P^\nu,\ \tilde{B}_\mu\equiv B_{\mu\nu}P^\nu,\ 
\gamma^2\equiv P_\nu P^\nu,\ \Omega\equiv E_{\mu\nu}P^\mu P^\nu.
\end{equation}
In terms of these quantities and 
$3+1$ quantities, the tensor $P_{\mu\nu}$ can be decomposed as:
\begin{eqnarray}
& &P_{\mu\nu}=(P^2+\gamma^2) B_{\mu\nu} +2Pn_{(\mu}\tilde{B}_{\nu)}
+2PE_{\tau(\mu}\epsilon^\tau_{\phantom{\tau}\nu)\lambda}P^\lambda\nonumber\\ 
& &-2P_{(\mu}\tilde{B}_{\nu)} 
+ l_{\mu\nu}\tilde{B}_{\sigma}P^\sigma-2 n_{(\mu}\tilde{E}_{|\tau}
\epsilon^\tau_{\phantom{\tau}\sigma|\nu)}P^\sigma. \label{P_decomp}
\end{eqnarray}
Consequently, the condition (\ref{type_Dc}) renders:
\begin{subequations}
\begin{eqnarray}
& &0=P_{\mu\nu}n^\mu n^\nu=\tilde{B}_{\nu}P^\nu, \label{short}\\
& &0=n^{\mu'}P_{\mu'\nu^\prime} h^{\nu^\prime}_{\phantom{\nu^\prime}\nu}
=-P\tilde{B}_{\mu}+ 
\tilde{E}_{\tau}\epsilon^\tau_{\phantom{\tau}\sigma\mu}P^\sigma, \label{wedge}\\
& &0=P_{\mu^\prime \nu^\prime} h^{\mu^\prime}_{\ \mu} h^{\nu^\prime}_{\ \nu} 
=(P^2+\gamma^2) B_{\mu\nu}
+2PE_{\tau(\mu}\epsilon^\tau_{\phantom{\tau}\nu)\lambda}P^\lambda\nonumber\\
& &\hspace{1cm}-2P_{(\mu}\tilde{B}_{\nu)} + h_{\mu\nu} \tilde{B}_{\lambda}P^\lambda.
\label{b-wedge}
\end{eqnarray}
\end{subequations}
\subsection{Decomposition of equation (\ref{type_Dd}')}

To find the decomposition of (\ref{type_Dd}') we need to decompose first 
$Q_{\mu\nu}$ and the vector field $\xi_\mu$. The latter is trivially 
decomposed in the form
$$
\xi_\mu=Yn_\mu+Y_\mu
$$
where $Y_\mu$ is spatial and $Y$ is a scalar. The decomposition of 
$Q_{\mu\nu}$ is far more involved  and it is not shown. Inserting all 
the decompositions just mentioned in (\ref{type_Dd}') we obtain the 
conditions
\begin{subequations}
\begin{eqnarray}
&&27 \alpha _{} \rho _{}^3Y_{}^2=\gamma _{}^2 \rho _{}-\Omega _{},\\ 
&&27\alpha _{} \rho _{}^3 Y_{} Y_{\nu }=\epsilon_{{\nu   }}^{{\ \lambda\sigma}} P_\lambda 
\tilde{B}_\sigma-P_{}\tilde{E}_{\nu }+\rho P_{}P_{\nu},\\
&&2 B_{{\sigma(\nu}}^{{  }} \epsilon _{{\mu)   }}^{{\ \sigma\lambda}}
P_\lambda P_{}-2P_{(\nu } \tilde{E}_{\mu) }
+E_{\mu \nu }^{{  }}(P_{}^2+\gamma _{}^2)\nonumber\\
&&-\rho _{}P_{\mu } P_{\nu } +h_{\mu \nu }^{{  }} 
((-P_{}^2+\gamma _{}^2) \rho _{}+\Omega _{})=
-27\alpha _{} \rho _{}^3Y_{\mu } Y_{\nu }.\nonumber\\
&&\label{electric_form} 
\end{eqnarray}
\end{subequations}

\subsection{Decomposition of equation (\ref{type_De})}
Finally, we note that the inequality (\ref{type_De}) can be written as
\begin{equation}
\alpha=\frac{1}{9\rho^2}(\gamma^2-P^2)+2\rho>0. \label{alpha_ineq}
\end{equation}

\section{Schwarzschild initial data: necessary conditions}
Now, we proceed to pull back the conditions (\ref{rule1}),
(\ref{rule2}), (\ref{wedge}), (\ref{b-wedge}), (\ref{electric_form}) 
 and the inequality (\ref{alpha_ineq}) to the manifold
$\mathcal{S}$. In such a way one obtains a first set of
\emph{necessary} conditions for a initial data set to be a
Schwarzschildean initial data set.

\begin{theorem} \label{first_necessary}
Let $(h_{ij}, K_{ij})$ be an initial data set and define from
it the following quantities
\begin{subequations}
\begin{eqnarray}
& & \rho=\left(\frac{1}{2} B_{i}^{\phantom{i}j}B^{il}E_{jl}
-\frac{1}{6} E_{ij}E_{l}^{\phantom{j}j}E^{il} \right)^{1/3},\label{let-rho}\\
& &P= -\frac{1}{2} E^{ij}
K_{ij}-\rho K\nonumber\\
& &\hspace{1cm}-\frac{1}{6\rho}\epsilon _{{\ \ i}}^{jk}
\left(E^{il} D_k B_{lj}+B^{il} D_k E_{lj}\right), \\
& &P_i=D_i\rho,\ \tilde{E}_i=E_{ij}P^j,\ \tilde{B}_i=B_{ij}P^j,\\ 
& &\gamma^2=P_{i}P^{i},\ \Omega=\tilde{E}_{i}P^i,\\
& &\alpha=\frac{\gamma^2-P^2}{9\rho^2}+2\rho,\ 
Y_{}^2=\frac{\gamma _{}^2 \rho _{}-\Omega _{}}{27 \alpha _{} \rho _{}^3},\\
& &
27\alpha _{} \rho _{}^3 Y_{}Y_{i }=\epsilon_{{i   }}^{{\ j k}} P_j 
\tilde{B}_k-P_{}\tilde{E}_{i }+\rho P_{}P_{i},\label{y_y}
\end{eqnarray}
\end{subequations}
where $E_{ij}$ and $B_{ij}$ are to be calculated from the initial
data $h_{ij}$ and $K_{ij}$ using the formulae (\ref{electric_idata})
and (\ref{magnetic_idata}).
Necessary conditions for an initial data set $(h_{ij},K_{ij})$,
satisfying the Einstein vacuum constraints, equations
(\ref{Hamiltonian}) and (\ref{Momentum}), on a manifold 
$\mathcal{S}$ to be a Schwarzschild initial data set are:
\begin{subequations}
\begin{eqnarray}
&&B_{ij}= -\frac{1}{\rho}
(B_{i\phantom{k}}^{\phantom{i}k} E_{kj}+B_{j\phantom{k}}^{\phantom{j}k}E_{ki}),\label{new_rule1}\\
&&E_{ij}= \frac{1}{\rho}(B_{i\phantom{k}}^{\phantom{i}k} B_{kj}
-E_{i\phantom{k}}^{\phantom{i}k} E_{kj})+2\rho h_{ij},\label{new_rule2} \\
&& \tilde{B}_{j} P^j =0, \label{new_short} \\
&& \tilde{E}_{k}\epsilon^k_{\phantom{k}li}P^l-P\tilde{B}_{i}=0, 
\label{new_wedge} \\
&&
(P^2+\gamma^2)B_{ij}
+2PE_{k(i}\epsilon^k_{\phantom{k}j)l}P^l-2P_{(i}\tilde{B}_{j)}=0,
\label{new_bwedge} \\
&& 2 B_{{l(i}}^{{  }} \epsilon _{{j)   }}^{{\ lk}}
P_k P_{}-2P_{(i } \tilde{E}_{j) }
+E_{j i }^{{  }}(P_{}^2+\gamma _{}^2)\nonumber\\
&&-\rho _{}P_{j } P_{i } +h_{j i }^{{  }} 
((-P_{}^2+\gamma _{}^2) \rho _{}+\Omega _{})=
-27\alpha _{} \rho _{}^3Y_{j } Y_{i },\nonumber\\
&&\label{new_electric_form}\\ 
&& \left(\frac{1}{2} B_{i}^{\phantom{i}j}B^{il}E_{jl}
-\frac{1}{6} E_{ij}E_{l}^{\phantom{j}j}E^{il} \right)^{1/3}\neq 0,
\label{new_rho}\\
&& \frac{1}{9\rho^2}(\gamma^2-P^2)+2\rho>0.\label{new_alphaineq} 
\end{eqnarray}
\end{subequations}
 \label{necessary-prop}
\end{theorem}

\section{Algebraic classification of Schwarzschild initial data}
\label{analysis}
Using the conditions found in theorem \ref{necessary-prop} we can
determine the possible types of Schwarzschild initial data one can
have, according to the possible admissible algebraic structure of the
tensors $E_{ij}$ and $B_{ij}$. In order to do this, we will make use
of adapted bases in certain subspaces of the tangent space of a point
$p\in \mathcal{M}$.  This subspace plays the role of the tangent space
of $\phi(\mathcal S$) at $p$.  We will establish the possible
\emph{canonical forms} for the matrices of the components of the
tensors $E_{ij}$ and $B_{ij}$ in these canonical bases. This is
achieved by means of algebraic computations involving the conditions
of theorem \ref{necessary-prop}.  Therefore the canonical forms
obtained here are only valid at $p$ and we should prove a posteriori
that initial data with $E_{ij}$ and $B_{ij}$ adopting a given
canonical form do in fact exist.  This is a clear consequence of the
fact that conditions supplied by theorem \ref{necessary-prop} are
only necessary.  More generally, one may construct initial data sets
such that the {\em algebraic types} deduced here for $E_{ij}$ and
$B_{ij}$ vary on $\mathcal S$. Indeed, some of the algebraic types
calculated here are attached to points lying in a specific region of
Schwarzschild-Kruskal spacetime and cannot be found outside these
regions.

An important result which we must bear in mind when developing the 
classification is the fact that $\rho$ and $\alpha$  never
vanish in the Kruskal-Schwarzschild spacetime, as can be checked by explicit
calculations in a coordinate system. Similarly one can show that 
the vector $\xi^\mu$ of (\ref{type_Dd}')
is spacelike only for points outside the static regions.

\subsection{Initial data with $Y=0$}
The condition $Y=0$ implies that the vector $\xi^\mu$ of (\ref{type_Dd}')
is spacelike 
so the canonical forms considered in this subsection only arise for 
submanifolds $\mathcal S$ whose points are in the exterior 
of the static regions.  
We consider two subcases defined by $P^i=0$ and $P^i\neq 0$. 

In the first case, 
the condition $P^i=0$ only occurs when $\nabla_a\rho$ is causal, which, 
consistently with the above remarks, is only possible for
 points outside the static regions as can be explicitly
checked by a calculation in a coordinate system. In this case 
the initial data hypersurface $\mathcal S$ belongs to the foliation defined by 
$\rho=const$. A not very long calculation using the conditions of theorem
\ref{necessary-prop} leads us to 
\begin{equation}
B_{ij}=0,\ E_{ij}=\rho h_{ij}-\frac{27\alpha\rho^3Y_i Y_j}{P^2},\ P\neq 0.
\end{equation}
If $P^i\neq 0$ then the canonical forms are somewhat more complex
\begin{eqnarray}
& &E_{ij}=\frac{-3\rho}{P^2-\gamma^2}P_iP_j
+\frac{\rho(-2\gamma^4+P^2(P^2+\gamma^2))}{(P^2-\gamma^2)^2}h_{ij}\nonumber\\
& &\hspace{1.3cm}-\frac{27\alpha\rho^2(P^2+\gamma^2)}{(P^2-\gamma^2)^2}Y_iY_j,\\
& & B_{ij}=\frac{27\alpha_{} \rho_{}^3
 P_k P_{} Y_l (\epsilon _{{i   }}^{{\ kl}} Y_{j}
+\epsilon _{{j  }}^{{\ kl}}Y_{i })}
{(P_{}^2-\gamma _{}^2)^2},
\end{eqnarray}
where $Y^j$ is orthogonal to $P^j$ and 
\begin{equation}
Y^jY_j=\frac{P^2-\gamma^2}{9\alpha\rho^2}.
\end{equation}
Of course these canonical forms are only valid 
for points with $P^2-\gamma^2\neq 0$ (points not intersecting the
event horizon).

\subsection{Initial data with $Y\neq 0$}
The condition $Y\neq 0$ implies that $P^i\neq 0$ and thus $\gamma\neq 0$.
Also note that if $Y\neq 0$ then $Y$ and $Y^j$ are fixed by the initial 
data by means of equations (\ref{y-1}) and (\ref{y-2}). 
We divide up these initial data in two cases: those with $\tilde{B}^iP=0$
and those with $\tilde{B}^iP\neq 0$.

\subsubsection{Case $P\tilde{B}^i=0$}
\label{pure-e}

We find next the form which conditions of theorem \ref{necessary-prop} take
when $P\tilde{B}^{i}=0$.
The conditions (\ref{new_rule1})-(\ref{new_bwedge}) in proposition
\ref{first_necessary} will be interpreted as defining a certain
\emph{canonical orthonormal basis} for our problem. In particular, if
$P\tilde{B}^{i}=0$, then condition (\ref{new_wedge}) reduces to
\begin{equation}
\tilde{E}_{k}\epsilon^k_{\phantom{k}li}P^l=0.
\end{equation}
The latter condition implies that $\tilde{E}^k$ has to be
proportional to $P^k$ ---that is, $P^k$ is an eigenvector of
$E_j^{\phantom{j}k}$,
\begin{equation}
E_j^{\phantom{j}k}P^j=\lambda P^k.
\label{eigenvalue}
\end{equation}
In view of the above, it is natural to consider an
orthonormal basis $\{(e_a)^i\}=\{(e_1)^i, (e_2)^i, (e_3)^i\}$, such that
\begin{equation}
P^i=\gamma (e_1)^i, \quad \tilde{E}^i=\Xi(e_1)^i,\ 
\Xi=\pm\sqrt{\tilde{E}_j\tilde{E}^j}.
\end{equation}
The remaining frame vectors $(e_2)^i$ and $(e_3)^i$ are left
undetermined. Now, the conditions (\ref{new_wedge}) and
(\ref{new_bwedge}) can be used to make further statements about the
the matrices  whose entries are the
components of the tensors $E_{ab}$ and $B_{ab}$ in the frame 
$\{(e_{\boldsymbol a})^{i}\}$. 
These matrices are defined by $E_{\boldsymbol a\boldsymbol b}
=E_{ij}
(e_{\boldsymbol a})^{i}(e_{\boldsymbol b})^{j}$, 
$B_{\boldsymbol a\boldsymbol b}=B_{ij}(e_{\boldsymbol a})^{i}
(e_{\boldsymbol b})^{j}$ .
One finds that
\begin{subequations}
\begin{eqnarray}
&& E_{12}=E_{13}=0,\ E_{11}=\frac{\Xi}{\gamma}\\
&&\hspace{-.8cm} B_{22}=\frac{2E_{23}P\gamma}{P^2+\gamma^2}=-B_{33},\ 
B_{23}=\frac{(E_{33}-E_{22})P\gamma}{P^2+\gamma^2},\\
&& B_{11}=B_{12}=0
\end{eqnarray}
\end{subequations}
From here we easily deduce $\tilde{B}^a=0$.
These expressions
 can be further restricted if we apply (\ref{new_electric_form})
---previously we need to use (\ref{y_y}) to find $Y$ and $Y^j$ in our chosen
basis, see (\ref{kid_candidate1}) below. 
As a result we find that all the components of the magnetic part are
zero ---purely electric initial data sets---
 whereas the electric part takes the form
\begin{subequations}
\begin{equation}
E_{23}=0,\ E_{22}=E_{33}=-\frac{\rho(\gamma^2-P^2)+\Omega}{P^2+\gamma^2},
\end{equation}
with
\begin{equation}
\Xi^2=\frac{4\gamma^2(\rho(\gamma^2-P^2)+\Omega)^2}
{(P^2+\gamma^2)^2}.
\end{equation}
\end{subequations}
Finally, contracting (\ref{new_rule2}) with 
$(e_{\boldsymbol a})^{i}(e_{\boldsymbol b})^{j}$ we get
\begin{equation}
E_{11}=-2\rho, \quad E_{22}=E_{33}=\rho.
\end{equation}
Therefore the electric part can be written in 
abstract index notation as
\begin{equation}
E_{ij}=\rho\left(h_{ij}-\frac{3}{\gamma^2}P_iP_j\right).
\label{a-eform}
\end{equation}
This last equation comprise the most general form for the electric
part which fulfills conditions (\ref{new_rule1})-(\ref{new_bwedge})
whenever $\tilde{B}^a=0$ (and as a consequence $B_{ij}=0$ as proven before).
It can be readily checked that condition (\ref{easy-1}) is
automatically satisfied by (\ref{a-eform}), 
and thus provides no further information. 

The class of purely electric initial data sets includes a number of
relevant examples. Time symmetric initial data sets ---that is, those
for which $K_{ij}=0$--- are tri\-vi\-ally contained in this class. 
Another example
of purely electric data sets is given by those having spherical
symmetry ---see example \ref{spheric-symmetric}.

Using the above results 
we can also supply canonical expressions for $Y$ and $Y^j$
\begin{equation}
Y=\frac{\gamma}{3\rho\sqrt{\alpha}},\ Y^i=\frac{P}{3\rho\sqrt{\alpha}}(e_1)^i.
\label{kid_candidate1}
\end{equation}
These expressions will be important later, for they give rise to {\em
  Killing Initial Data candidates} which, under certain conditions,
enable us to show that initial data whose electric and magnetic part
fulfill the conditions obtained in this subsection are actually
Schwarzschildean initial data.

\begin{example}[Spherically symmetric data]
\label{spheric-symmetric}
 \em In the so-called \emph{proper distance
gauge} spherically symmetric first and second fundamental forms look like
\begin{subequations}
\begin{eqnarray}
&& h_{ij}dx^idx^j=dl^2+R^2(l)(d\theta^2+\sin^2\theta d\varphi^2), 
\label{spheric-h}\\ 
&&\hspace{-1cm}
 K_{ij}dx^idx^j=K_L(l)dl^2+K_R(l)(d\theta^2+\sin^2\theta d\varphi^2)
\label{spheric-k}
\end{eqnarray}
\end{subequations}
where $u^i$ is the outward-pointing normal to the spheres with
constant $R$, $l$ is a coordinate measuring the proper distance
between spheres of fixed radius and $K_L(l)$, $K_R(l)$ are 
scalar functions---see \cite{OMuRos06}.
The Hamiltonian and momentum
constraints for these data are given respectively by 
\begin{subequations}
\begin{eqnarray}
&&\hspace{-1cm}
 K_R(K_R+2K_L)-\frac{1}{R^2}(2RR^{\prime\prime}+R^{\prime 2}-1)=0, \\
&& K^\prime_R+\frac{R^\prime}{R}(K_R-K_L)=0,
\end{eqnarray}
\end{subequations}
where the prime denotes the derivative with respect to $l$. If these equations hold 
then it can be explicitly checked that $B_{ij}=0$. As a result,
conditions (\ref{new_rule1})-(\ref{new_bwedge}) are automatically
satisfied.  This can be regarded as a 3-dimensional manifestation of
Birkhoff's theorem because the development of the data given by
(\ref{spheric-h}), (\ref{spheric-k}) is a vacuum spherically symmetric
spacetime and thus, by Birkhoff's theorem, a patch of Schwarzschild
spacetime. Therefore the data (\ref{spheric-h}), (\ref{spheric-k}) are
automatically Schwarzschild initial data if the vacuum constraints
hold.  It is noteworthy to point out that the approach in
\cite{OMuRos06} is somehow different in as much that they do not
assume \emph{a priori} that the constraint equations are satisfied.
\end{example}

\subsection{Initial data sets with $\tilde{B}^i\neq 0$}
In this case, the most general one of our analysis, $P^i$ ceases to be
an eigenvalue of $E_i^{\phantom{i}j}$, however, it will be still an
element of our canonical basis. We write as before, 
$P^i=\gamma (e_1)^i$ so that again
$B_{11}=(e_1)^i(e_1)^jB_{ij}=0$. Now, the condition (\ref{new_short})
suggest to consider a second frame vector $(e_2)^i$ parallel to
$\tilde{B}^i$. The latter choice fixes automatically
$(e_3)^i$, which has to be parallel to $\epsilon^i_{\phantom{i}jk}P^j
\tilde{B}^k$. In what follows, we will write
\begin{equation}
\tilde{B}^{i}=\beta(e_2)^i, \quad \beta=\pm\sqrt{\tilde{B}_k\tilde{B}^k}.
\end{equation}
Using the frame $\{ (e_{\boldsymbol a})^i\}$ thus constructed, one can use the
conditions 
 (\ref{new_wedge}) and (\ref{new_bwedge})
to express the components 
$B_{\boldsymbol a\boldsymbol b}
=B_{ij}(e_{\boldsymbol a})^i(e_{\boldsymbol b})^j$ in terms of
$E_{\boldsymbol a\boldsymbol b}
=E_{ij}(e_{\boldsymbol a})^i(e_{\boldsymbol b})^j$, 
and the scalars $\beta$, $\gamma$ and
$P$. More precisely, one has that
\begin{equation}
E_{12}=0,\ E_{13}=\frac{P\beta}{\gamma^2}
\end{equation}
and
\begin{subequations}
\begin{eqnarray}
&& B_{12}=\frac{\gamma}{P}E_{13}=\frac{\beta}{\gamma}, \\
&& B_{13}=0, \\
&& B_{22}=\frac{2\gamma PE_{23}}{P^2+\gamma^2}, \\
&& B_{33}=-\frac{2\gamma PE_{23}}{P^2+\gamma^2}, \\
&& B_{23}=\frac{\gamma P (E_{33}-E_{22})}{P^2+\gamma^2}.
\end{eqnarray}
\end{subequations}
As in subsection \ref{pure-e}, we now resort to conditions 
(\ref{new_electric_form}), (\ref{new_rule1}), (\ref{new_rule2})
 to obtain further restrictions on the
components of $E_{ij}$ and $B_{ij}$. We spare the reader 
from the intermediate long calculations and  just provide the 
final result for the components
of the electric and magnetic parts which are
\begin{widetext}
\begin{subequations}
\begin{eqnarray}
&&(E_{\boldsymbol a\boldsymbol b})=
\begin{pmatrix}
-\frac{\Omega}{\gamma^2} & 0 & 
\frac{P}{\gamma^2}\sqrt{\frac{(\Omega-\gamma^2\rho)
(2\gamma^2\rho+\Omega)}{\gamma^2-P^2}} \\
0 & \frac{\rho(P^2+\gamma^2)+\Omega}{-\gamma^2+P^2} & 0 \\
\frac{P}{\gamma^2}\sqrt{\frac{(\Omega-\gamma^2\rho)
(2\gamma^2\rho+\Omega)}{\gamma^2-P^2}} & 
0 & -\frac{\gamma^4\rho+P^2(\gamma^2\rho+\Omega)}{\gamma^2(P^2-\gamma^2)}
\end{pmatrix},\\  
&&(B_{\boldsymbol a\boldsymbol b})=
\begin{pmatrix}
0 &\frac{P}{\gamma^2}\sqrt{\frac{(\Omega-\gamma^2\rho)
(2\gamma^2\rho+\Omega)}{\gamma^2-P^2}} & 0 \\
\frac{P}{\gamma^2}\sqrt{\frac{(\Omega-\gamma^2\rho)
(2\gamma^2\rho+\Omega)}{\gamma^2-P^2}} & 0 & 
\frac{P(2\gamma^2\rho+\Omega)}{\gamma(P^2-\gamma^2)} \\
0 &\frac{P(2\gamma^2\rho+\Omega)}{\gamma(P^2-\gamma^2)} & 0
\end{pmatrix}.\ \ \ \ \label{full-1}
\end{eqnarray}
\end{subequations}
\end{widetext}
These canonical forms assume the conditions 
$P^2-\gamma^2\neq 0$ so they only hold 
for points which are not in the event horizon.
Also, as in previous case, canonical expressions for  $Y$ and $Y^j$ can be 
obtained
\begin{eqnarray}
&&Y=\frac{\sqrt{\gamma^2\rho-\Omega}}{3\rho^{3/2}\sqrt{3\alpha}},\\
&&Y^{i }=(e_{ 1})^{{i  }}\frac{P_{} 
\sqrt{\gamma _{}^2 \rho _{}-\Omega _{}}}
{3 \gamma _{} \rho _{}^{3/2}\sqrt{3\alpha}} \nonumber\\
&&\hspace{1cm}-(e_{3})^{{i }}\frac{\sqrt{(P_{}^2-\gamma _{}^2) 
(2 \gamma _{}^2 \rho _{}+\Omega _{})}}
{3\gamma _{} \rho _{}^{3/2}\sqrt{3\alpha}}.
\end{eqnarray}

\section{Sufficient conditions for Schwarzschild initial data}
\label{sufficient}
In order to discuss whether or not the conditions given in proposition
\ref{first_necessary} are also sufficient conditions for a pair
$(h_{ij},K_{ij})$ of symmetric tensors on a 3-dimensional manifold
$\mathcal{S}$ which satisfy the Einstein constraint equations to be a
Schwarzschild initial data set, one is confronted with the issue of the
propagation of the conditions (\ref{new_rule1})-(\ref{new_alphaineq}).
One has to show that the development of initial data satisfying our
conditions ---or an extended set of them--- is a subset of
Schwarzschild spacetime. More precisely, assume that on a given
3-dimensional manifold $\mathcal{S}$, the initial data
$(h_{ij},K_{ij})$ satisfies the conditions presented in
theorem \ref{first_necessary}, and set
$S_0\equiv\phi(\mathcal{S})$ with 
$\phi:{\mathcal S}\rightarrow{\mathcal M}$ an isometric embedding. 
Then if one is able to prove ---under, perhaps, some
extra conditions--- that conditions
(\ref{type_Da})-(\ref{type_De}) given in theorem \ref{thm_FerSae} are
also satisfied in an open subset $\mathcal{N}$ 
of the development which contains $S_0$, 
it follows that $\mathcal{N}$ would be a portion of the
Schwarzschild spacetime. Note that due to the uniqueness of the
maximal development of the initial data ---see e.g. \cite{ChoGer69}---
this would prove that the maximal development is also a patch of 
Schwarzschild spacetime. A similar idea could be applied if
one is considering a subset $\mathcal{U}\subset \mathcal{S}$.

To address the problem explained in
previous paragraph we will require that the development of our initial
data ---or at least a subset $\mathcal{U}\subset \mathcal{S}$
thereof--- possesses a timelike Killing vector $\xi^\mu$. In that
case, denoting by $\phi_t$ the local flow generated by $\xi^\mu$ then
one may seek conditions under which
(\ref{new_rule1})-(\ref{new_bwedge}) hold in the family of
hypersurfaces $\phi_t(S_0)$, $t\in (-\epsilon,\epsilon)$ which
 would imply that conditions
(\ref{type_Da})-(\ref{type_De}) of theorem \ref{thm_FerSae} are
also satisfied in an open subset $\mathcal{N}$ of  the development. 

In what follows, it is assumed that the initial data set $(h_{ij},K_{ij})$
besides fulfilling the conditions (\ref{new_rule1})-(\ref{new_alphaineq}),
is also sufficiently regular and possesses the required asymptotic decay so
that at least an open set, $\mathcal{N}$, of development exists.

\subsection{Killing initial data sets}
A Killing initial data set ---a \emph{KID}--- associated to the initial data 
$(\mathcal{S},h_{ij},K_{ij})$ is a pair
$(Y,Y^i)$ consisting of a scalar $Y$ and a vector $Y^i$ defined on
$\mathcal{S}$ satisfying the following system of linear partial differential
equations ---\emph{the KID equations}---
\begin{subequations}
\begin{eqnarray}
&& D_{(i}Y_{j)}-Y K_{ij}=0, \label{kid-1}\\ 
&&\hspace{-1cm} D_iD_jY-\mathcal{L}_{Y^l} K_{ij} =Y(r_{ij}+
KK_{ij}-2K_{il}K^l_{\phantom{l}j})\label{kid-2}.
\end{eqnarray}
\end{subequations} 
For any KID there exists a Killing vector $\xi^\mu$ in the de\-ve\-lop\-ment 
of $(\mathcal{S},h_{ij}, K_{ij})$ whose $3+1$ 
decomposition when pull-backed to $\mathcal{S}$ is precisely the KID
\cite{BeiChr97b,Coll77,Mon75}. More precisely, if we consider any foliation
of the development with $\phi(\mathcal S)$ as one of its leaves then 
the $3+1$ decomposition of $\xi^\mu$ reads
\begin{equation}
\xi^\mu=Yn^\mu+Y^\mu,
\label{kid-split}
\end{equation}
where $n^\mu$ is the normal vector to the leaves. 
In this case $\phi^*Y$ and $Y_i=\phi^*Y_\mu$
satisfy the equations (\ref{kid-1}) and (\ref{kid-2}) on $\mathcal S$.

Determining whether an initial data set possesses a KID is in general
a non-trivial endeavour which entails finding solutions of the system
(\ref{kid-1})-(\ref{kid-2}). Fortunately, theorem \ref{thm_FerSae}
and theorem \ref{necessary-prop} enables us to find an Ansatz to
solve these equations:
\begin{subequations}
\begin{eqnarray}
&&Y=\sqrt{\frac{\gamma^2\rho-\Omega}
{27\alpha\rho^3}},\label{y-1}\\
&&Y^i=\frac{\rho PP^i-\tilde{E}^i P+\tilde{B}^k P_{l }
\epsilon^{il}_{\phantom{il}k} }{\sqrt{27\alpha\rho^3(\gamma^2\rho-\Omega)}}.
\label{y-2}
\end{eqnarray}
\end{subequations}

The pair $(Y,Y^i)$ defined by the above equations will be called a
{\em KID candidate}. Clearly a KID candidate can be always constructed
from any initial data set with $\alpha\neq 0$, $\rho\neq 0$ and
$\gamma^2\rho-\Omega>0$.  It is plausible to conjecture that for an
initial data set satisfying the conditions given in theorem
\ref{first_necessary}, the KID candidate satisfies the KID equations
(\ref{kid-1}) and (\ref{kid-2}).  A possible approach to how this
conjecture could be attacked is discussed in the appendix for the
simpler case of time symmetric data. In view of the computational
difficulties in proving this assertion, we shall adopt here a
pragmatic perspective and require that in addition to satisfying the
conditions in theorem \ref{first_necessary} the initial data set
$(h_{ij},K_{ij})$ is such that the KID candidate solves the KID
equations, so that at least a portion of the resulting spacetime
possesses a timelike Killing vector ---namely that portion in the
domain of dependence of the subset $\mathcal{U}\subset\mathcal{S}$
where $Y^iY_i-Y^2<0$ (timelike KID). 

A simpler KID candidate can be found from (\ref{y-1}) and (\ref{y-2})
if we eliminate $\alpha$ using (\ref{mass}). Since $m$, given by
(\ref{mass}) has to be a constant for Schwarzschildean initial data
sets, it can be also removed from the expression for the KID yielding
\begin{equation}
Y=\frac{\sqrt{\gamma^2\rho-\Omega}}
{\rho^{11/6}},\ 
Y^i=\frac{\rho PP^i-\tilde{E}^i P+\tilde{B}^k P_{l }
\epsilon^{il}_{\phantom{il}k} }{\rho^{11/6}\sqrt{\gamma^2\rho-\Omega}}.
\label{kid-simple}
\end{equation}

\subsection{Main result}
The following
theorem can be regarded as a 
converse of theorem (\ref{necessary-prop}).
\begin{theorem}
  Let $(\mathcal{S},h_{ij},K_{ij})$ be an initial data set satisfying
  the vacuum Einstein constraint equations (\ref{Hamiltonian}) and
  (\ref{Momentum}) and in addition the conditions (\ref{new_rule1}),
  (\ref{new_rule2}), (\ref{new_short}), (\ref{new_wedge}),
  (\ref{new_bwedge}), (\ref{new_electric_form}), (\ref{new_rho}) and
  (\ref{new_alphaineq}) of theorem \ref{necessary-prop}. Further,
  let the quantities
\begin{eqnarray}
&&Y=\frac{\sqrt{\gamma^2\rho-\Omega}}
{\rho^{11/6}},\ \gamma^2\rho-\Omega>0,\label{main-1} \\
&&Y^i=\frac{\rho PP^i-\tilde{E}^i P+\tilde{B}^k P_{l }
\epsilon^{il}_{\phantom{il}k} }
{\rho^{11/6}\sqrt{\gamma^2\rho-\Omega}},\label{main-2}
\end{eqnarray}
solve the KID equations (\ref{kid-1}) and (\ref{kid-2}) in an open subset
$\mathcal{U}\subset\mathcal{S}$ with
\begin{equation}
 Y^i Y_i-Y^2<0,
\label{kid-inequality}
\end{equation}
then $(\mathcal{U},h_{ij},K_{ij})$ is isometric to a region of an
initial data set $(\mathcal{S}^\prime,h^\prime_{ij},K^\prime_{ij})$
for the Schwarzschild spacetime.
\label{main_result}
\end{theorem}

\noindent
{\it Proof :}\hspace{3mm}
According to standard results, the development $\mathcal{D}(\phi(\mathcal S))$ 
is a vacuum solution of Einstein equations so we may define its Weyl tensor 
$C_{\mu\nu\lambda\sigma}$ in the standard fashion. 
Conditions (\ref{kid-1})-(\ref{kid-2}) imply that 
a Killing vector $\xi^\mu$ exists in 
$\mathcal{N}\cap\mathcal{D}(\phi(\mathcal{U}))$ where 
$\mathcal N$ is an open set containing $\phi(\mathcal U)$. 
Let $n^\mu$ be the normal vector of a foliation of $\mathcal N$ 
adapted to $\phi(\mathcal U)$ and define the $3+1$ splitting of 
$\xi^\mu$ as in (\ref{kid-split}).
By continuity, $\mathcal N$ can be chosen 
in such a way that $Y^\mu Y_\mu-Y^2<0$ on 
$\mathcal{N}\cap \mathcal{D}(\phi(\mathcal{U}))$ 
if $Y^iY_i-Y^2<0$ on $\mathcal{U}$ from which we deduce that 
$\xi^\mu$ is timelike in $\mathcal{N}\cap \mathcal{D}(\phi(\mathcal{U}))$.

Now, 
$\mathcal{L}_\xi C_{\mu\nu\sigma\lambda}=0$ and any concomitant constructed 
exclusively from the Weyl tensor $C_{\mu\nu\sigma\lambda}$ will satisfy a
 similar property. Therefore we deduce
\begin{eqnarray}
&&\mathcal{L}_\xi ((S\star S)_{\mu\nu\sigma\lambda}-S_{\mu\nu\sigma\lambda})=0,\ 
\mathcal{L}_\xi P_{\mu\nu}=0,\nonumber\\ 
&&\mathcal{L}_\xi(Q_{\mu\nu}+9\alpha\rho^2\xi_\mu\xi_\nu)=0.\label{system-1}
\end{eqnarray}
Also the conditions (\ref{new_rule1})-(\ref{new_alphaineq}) entail
\begin{equation}
((S\star S)_{\mu\nu\sigma\lambda}-S_{\mu\nu\sigma\lambda})
|_{\phi(\mathcal U)}=0,\ 
P_{\mu\nu}|_{\phi(\mathcal U)}=0,
\label{system-2}
\end{equation}
 and (\ref{main-1})-(\ref{main-2}) imply
\begin{equation}
(Q_{\mu\nu}+9\alpha\rho^2\xi_\mu\xi_\nu)|_{\phi(\mathcal U)}=0,
\label{system-3}
\end{equation}
Equations (\ref{system-1})-(\ref{system-3}) can be regarded as a Cauchy initial
value problem for a first order Linear system of PDE's if we take
$(S\star S)_{\mu\nu\sigma\lambda}-S_{\mu\nu\sigma\lambda}$,
$P_{\mu\nu}$ and $Q_{\mu\nu}+9\alpha\rho^2\xi_\mu\xi_\nu$ 
as the unknowns. Therefore the uniqueness of its 
solutions implies 
\begin{eqnarray}
&&(S\star S)_{\mu\nu\sigma\lambda}-S_{\mu\nu\sigma\lambda}=0,\ 
P_{\mu\nu}=0,\nonumber\\
&& Q_{\mu\nu}+9\alpha\rho^2\xi_\mu\xi_\nu=0
\end{eqnarray}
in $\mathcal{N}'\cap\mathcal{D}(\phi(\mathcal{U}))$, where
 $\mathcal{N}'$ is an open set containing $\phi(\mathcal U)$.
Similarly, we can construct the system
\begin{equation}
\mathcal{L}_{\xi}\rho=0, \quad \mathcal{L}_{\xi}\alpha=0,
\end{equation}
whose initial data, by (\ref{new_rho}) and (\ref{new_alphaineq}), have 
the properties 
\begin{equation}
\rho|_{\phi(\mathcal{U})}\neq 0,\  \alpha|_{\phi(\mathcal{U})}>0
\end{equation}
The properties of the initial data clearly guarantee that 
the solutions of the system are such that $\rho\neq 0$ and $\alpha>0$ in
$\mathcal{N}''\cap\mathcal{D}(\phi(\mathcal{U}))$ where $\mathcal{N}''$ 
is an open set with similar properties as $\mathcal N'$. 
Thus, we conclude that conditions of theorem \ref{thm_FerSae}
hold on $\mathcal{N}\cap\mathcal{N}'\cap\mathcal{N}'' 
\cap\mathcal{D}(\phi(\mathcal{U}))$
and accordingly this set is isometric to a portion of the
Schwarzschild spacetime. $\blacksquare$  

\medskip
\noindent
\textbf{Remarks.}  
If a vacuum initial data set
$(\mathcal{S},h_{ij},K_{ij})$ satisfies just conditions 
(\ref{new_rule1})-(\ref{new_alphaineq})
in some open region of $\mathcal{S}$, then it is
possible in principle 
for the  initial data set to be Schwarzschildean. One would be
forced, for example, to show the existence of a timelike KID by other
means. However, it is tempting to conjecture that if a vacuum initial data
set satisfying (\ref{new_rule1})-(\ref{new_alphaineq}) has a timelike
KID, it has to be of the form given by (\ref{main-1})-(\ref{main-2}).

\section{Conclusions}
In this work we have formulated necessary and sufficient conditions
for initial data set to be a Schwarzschild initial data set. The set
of necessary conditions and the set of sufficient conditions are not
the same but they have common features. The difference between both
sets of conditions is the presence of the KID equations in the set of
sufficient conditions.  Ideally, one would expect that the necessary
conditions obtained out of the invariant characterisation of
\cite{FerSae98} should imply that the the KID candidate given by
(\ref{y-1}) and (\ref{y-2}) actually solves KID equations.  The proof
of this conjecture in the general case remains unknown.  Therefore we
have adopted in this work the pragmatic and algorithmic perspective of
requiring as part of our sufficiency conditions that the KID candidate
is \emph{actually} a timelike KID, at least for an open region $\mathcal{U}$ of
the initial hypersurface. Due to the correspondence between KIDs and
Killing vectors in the development, this last hypothesis implies the
existence of a timelike Killing vector in the development of the region
$\mathcal{U}$. This is a standard hypothesis in some characterisations
of exact solutions ---see e.g. \cite{Mar99,Mar00,Sim84}.

The results presented in this work may be of use for numerical
relativists working in the simulation of black hole spacetimes. In
particular, one could wonder whether the formulae (\ref{y-1}) and
(\ref{y-2}) which can be calculated at every time step of a numerical
simulation can be used as some sort of \emph{symmetry seeking gauge}
---i.e. lapse and shift--- in evolutions where the late stages are
describable by the Schwarzschild solution ---like in case of the head-on
collision of black holes. Further, one could ask if out of the
necessary conditions presented in this work it is possible to
construct a measure telling how a given pair of intrinsic metric and
extrinsic curvature, $(h_{ij},K_{ij})$, differs from a Schwarzschild
initial data. Such a construct should be of value to analyse the
validity of perturbation schemes. These, and related issues will be
discussed elsewhere.

\acknowledgments
We would like to thank JMM Senovilla, R Beig C Lechner and J Ib\'a\~nez
for useful discussions.  
Also we thank JM Mart\'{\i}n-Garc\'{\i}a for making available 
to us an advanced version of his computer program {\em xTensor} 
prior to its release and for his technical aid with the program. 
AGP wishes to thank the 
School of Mathematical Sciences of Queen Mary College in London, where
most of this work was carried out, for 
hospitality.
JAVK's research is funded by an EPSRC Advanced Research
Fellowship. AGP is supported by research grants FIS2004-01626 of the Spanish
MEC and no. 9/UPV 00172.310-14456/2002 of Universidad del Pa\'{\i}s Vasco 
 in Spain.

\appendix

\section{Practical example: time-symmetric initial data} \label{kid_ts}
In this appendix we explain how to use 
theorem \ref{main_result} to the case of time symmetric 
initial data sets. 
In such case one has $K_{ij}=0$ which entails
\begin{equation}
E_{ij}=r_{ij}, \quad B_{ij}=0,\quad P=0,\ 
\end{equation}
and the Hamiltonian constraint reduces to $r=0$. Also we can take advantage 
of the results of subsection \ref{pure-e} and specially of the 
expressions for the electric part written in (\ref{a-eform}). 
With these assumptions one can easily show that 
conditions (\ref{new_rule1})-(\ref{new_electric_form}) of theorem 
\ref{necessary-prop} hold. 
The only nontrivial conditions which must be checked are
(\ref{new_rho})-(\ref{new_alphaineq}) and the
 fact that the quantities defined by (\ref{main-1}) and (\ref{main-2}) solve
the KID conditions (\ref{kid-1})-(\ref{kid-2}) and satisfy
(\ref{kid-inequality}).

The conditions $P=0$, $B_{ij}=0$ imply that the 
shift $Y^i$ of a KID candidate vanishes.
 Consequently the first KID equation (\ref{kid-1}) is satisfied
automatically while (\ref{kid-2}) reduces to
\begin{equation}
D_iD_jY=YE_{ij}.
\label{kid2-symmetric}
\end{equation}
For data with $B_{ij}=0$ we have $\Omega=-2\gamma^2\rho$ and thus
 equation (\ref{main-1}) becomes
\begin{equation}
Y=\frac{\gamma}{\rho^{4/3}}.\ 
\label{y-symmetric}
\end{equation}
Clearly condition (\ref{kid-inequality}) is fulfilled.
 Replacing (\ref{y-symmetric}) in (\ref{kid2-symmetric}) 
we obtain after some algebra
\begin{eqnarray}
&&-28\gamma P_i P_j+9{E}_{{ij}}^{{  }} \gamma{\rho}^2\nonumber\\
&&\hspace{-1.5cm}+12 {\rho}(P_j D_i\gamma+P_iD _j\gamma)+12\gamma\rho D_jP_i
-9 {\rho}^2 D_jD _i\gamma=0.
\label{kid2-condition}
\end{eqnarray}
Now, if we are to prove that certain 
time-symmetric initial data 
are actually Schwarzschild initial data then
 we need to show that (\ref{kid2-condition}) becomes an identity.
To illustrate how this works in practise, 
we shall consider an initial data 
set such that the metric $h_{ij}$ is given by
\begin{eqnarray}
&& h_{ij}dx^idx^j=dx^2+\frac{1}{F^2(x,y,z)}(dy^2+dz^2),\nonumber\\ 
&& F(x,y,z)>0,\ F(x,y,z)\in C^{\infty}(\mathcal S).
\label{data-set}
\end{eqnarray}
The coordinates $\{x,y,z\}$ cover the whole manifold $\mathcal S$.
We make no assumptions at this stage about the topology of
$\mathcal S$. The coordinate $x$ is adapted to $P_i$
in such a way that $P_idx^i=\gamma dx$.
 The aim is to find for which function $F$ the initial data 
$({\mathcal S},h_{ij},K_{ij})$ 
is a time-symmetric Schwarzschild initial data set.

Following our discussion in subsection \ref{pure-e}, we set up
an orthonormal frame to carry out the calculations.
The elements of the orthonormal frame are defined by
\begin{equation}
\frac{P^i\partial_i}{\gamma}=(e_1)^i\partial_i=\partial_x,
\ (e_2)^i\partial_i=F\partial_y,\
(e_3)^i\partial_i=F\partial_z.
\end{equation}
Note that, since the vector fields $(e_2)^i$ and $(e_3)^i$ are orthogonal
to the integrable 1-form $P_idx^i$, 
they give rise to an integrable 2-dimensional
distribution.
The coordinates $(y,z)$ parametrise the integral submanifolds
 of the distribution spanned by $(e_2)^i$ and $(e_3)^i$.
Also, according to (\ref{a-eform}) any tangent vector to
any of these submanifolds is an eigenvector of $E^i_{\ j}$
with eigenvalue $\rho$ 
which means that the basis vectors $(e_2)^i$ and $(e_3)^i$ span the
eigenspace associated to the repeated eigenvalue $\rho$. 
Thus from (\ref{a-eform}) we deduce
\begin{equation}
E_{ij}=-2\rho(e^1)_i(e^1)_j+\rho(e^2)_i(e^2)_j+\rho(e^3)_i(e^3)_j.
\label{electric-frame}
\end{equation}
The equation $P_i=D_i\rho$ implies the relations 
$\partial_y\rho=\partial_z\rho=0$, 
$\partial_x\rho=\gamma$ to be used later.
These relations mean that in our adapted coordinates
both $\rho$ and $\gamma$ are functions of $x$ only.
Now, if we denote $\partial_x\rho=\rho^\prime$, 
$\partial_x\gamma=\gamma^\prime$ then
using (\ref{y-symmetric}) we get
\begin{equation}
Y=\rho^{-4/3}\gamma=\rho^{-4/3}\rho^\prime.
\end{equation}

Next we show that (\ref{kid2-condition}) is indeed an identity when expressed
in the adapted frame constructed previously. To that end we need
to work out the Ricci rotation coefficients associated to this frame
and from them the Riemann tensor and
the Ricci tensor. Our conventions for the Ricci rotation coefficients,
$\omega^{\boldsymbol a}_{\ \boldsymbol b\boldsymbol c}$,
commutation coefficients 
$c^{\boldsymbol a}_{\ \boldsymbol b\boldsymbol c}$ and
Riemann tensor components 
$r^{\boldsymbol a}_{\ \boldsymbol b\boldsymbol c\boldsymbol d}$ are
\begin{eqnarray}
&&(e_{\boldsymbol b})^i\nabla_i(e_{\boldsymbol c})^j=
\omega^{\boldsymbol d}_{\ \boldsymbol b\boldsymbol c}(e_{\boldsymbol d})^j,\ 
c^{\boldsymbol a}_{\ \boldsymbol b\boldsymbol c}=
\omega^{\boldsymbol a}_{\ \boldsymbol b\boldsymbol c}
-\omega^{\boldsymbol a}_{\ \boldsymbol c\boldsymbol b},\\
&& r^{\boldsymbol c}_{\ \boldsymbol f\boldsymbol a\boldsymbol b}=
\omega^{\boldsymbol c}_{\ \boldsymbol a\boldsymbol d}
\omega^{\boldsymbol d}_{\ \boldsymbol b\boldsymbol f}- 
\omega^{\boldsymbol c}_{\ \boldsymbol b\boldsymbol d}
\omega^{\boldsymbol d}_{\ \boldsymbol a\boldsymbol f} \nonumber\\
&&\hspace{1cm}+\Delta_{\boldsymbol a}\omega^{\boldsymbol c}_{\ \boldsymbol b\boldsymbol f}-
\Delta_{\boldsymbol b}\omega^{\boldsymbol c}_{\ \boldsymbol a\boldsymbol f}-
\omega^{\boldsymbol c}_{\ \boldsymbol d\boldsymbol f}
c^{\boldsymbol d}_{\ \boldsymbol a\boldsymbol b},
\end{eqnarray}
where $\Delta_{\boldsymbol a}$ is the directional derivative associated 
to the vector field $(e_{\boldsymbol a})^i$. 
In addition, we have the metric condition which in an orthonormal 
frame takes the form
\begin{equation}
\omega^{\boldsymbol a}_{\ \boldsymbol b\boldsymbol c}
-\omega^{\boldsymbol c}_{\ \boldsymbol b\boldsymbol a}=0.
\end{equation}
Evaluating the commutators of the directional derivatives
on the coordinates one finds that the only nonvanishing 
Ricci rotation coefficients are
\begin{equation}
\omega^{1}_{\ 33}
=\omega^{1}_{\ 22}=\frac{\partial_xF}{F},\ \omega^{2}_{\ 33}=\partial_yF,\
\omega^{2}_{\ 23}=-\partial_zF.
\end{equation}
With this information we can work out the 
components of the Ricci tensor in our orthonormal 
frame. Using the relation $r_{ij}=E_{ij}$ and (\ref{electric-frame}) we get
\begin{eqnarray}
&&-\frac{2 (\partial_xF_{})^2}{F_{}^2}+\frac{\partial^2_xF_{}}{F_{}}
=- {\rho}_{},\
-\frac{\partial_yF_{} \partial_xF_{}}{F_{}}+\partial^2_{xy}F_{}=0,\nonumber\\
&&-\frac{\partial_zF_{} \partial_xF_{}}{F_{}}+\partial^2_{xz}F_{}=0,
\label{riemann-1}\\
&& -(\partial_zF_{})^2+F_{} \partial^2_zF_{}-(\partial_yF_{})^2+F_{}
\partial^2_yF_{}-\nonumber\\
&&-\frac{3 (\partial_xF_{})^2}{F_{}^2}
+\frac{\partial^2_xF_{}}{F_{}}={\rho}_{},\label{riemann-2}
\end{eqnarray}
and the Hamiltonian constraint $r=0$ becomes
\begin{equation}
-(\partial_yF)^2-(\partial_zF)^2+F(\partial^2_yF+\partial^2_zF)
+\frac{2\partial^2_xF}{F}-\frac{5(\partial_xF)^2}{F^2}=0.
\label{hamiltonian}
\end{equation}
Later we will show how (\ref{riemann-1})-(\ref{hamiltonian}) can be solved.
Before doing that let us prove that (\ref{kid2-condition}) 
is an identity for our initial data. In our adapted frame 
(\ref{kid2-condition}) takes the form 
\begin{eqnarray}
&&(e^{ 2})_{i}(e^{ 2})_{j}\left(9\gamma{\rho}_{}^3-\frac{12 \gamma^2{\rho}_{} 
\partial_xF_{}}{F_{}}+\frac{9\alpha'{\rho}_{}^2 \partial_xF_{}}{F_{}}\right)
+\nonumber\\
&&+(e^{ 3})_{i}(e^{3})_{j}\left(9\gamma{\rho}_{}^3-\frac{12 \gamma^2{\rho}_{} 
\partial_xF_{}}{F_{}}+\frac{9\alpha'{\rho}_{}^2 \partial_xF_{}}{F_{}}\right)
+\nonumber\\
&&\hspace{-1cm}+(e^{1})_{{i}}(e^{ 1})_{{j}}(-28 \gamma^3-18\gamma{\rho}_{}^3
+36 {\rho}_{}\gamma\gamma' _{}
-9 {\rho}_{}^2 \gamma'')=0\label{kid2-expanded}.
\end{eqnarray}
We must show that all the expressions in brackets are zero if
(\ref{riemann-1}) and (\ref{riemann-2}) hold.
Our strategy will be to find relations for
$\gamma$, $\gamma'$, $\gamma''$
and then plug these relations into (\ref{kid2-expanded}).
All these derivatives can be calculated
by differentiating (\ref{riemann-1})-(\ref{riemann-2}) and
using $\rho'=\gamma$ but
in order to avoid long and messy calculations we will follow
an alternative procedure. We start with the second Bianchi identity
which in our case takes the simpler form
\begin{equation}
D^iE_{ij}=0.
\end{equation}
Replacing the electric part by its expression in
terms of the adapted frame, eq. (\ref{electric-frame}), we
find
\begin{equation}
2\rho'-3\rho(\omega^{1}_{\ 22}+\omega^{1}_{\ 33})=0, \label{rho_prime}
\end{equation}
from which, replacing the rotation coefficients
\begin{equation}
\rho'=\frac{3\rho\partial_xF}{F}=\gamma.
\end{equation}
These equations enable us to calculate $\gamma'$ and $\gamma''$ by
just differentiating. Note that in equation
(\ref{kid2-expanded}) there are no partial derivatives of $F$ of order greater
than one. Therefore, in order to keep our expressions for $\gamma'$ and
$\gamma''$ with partial derivatives of $F$ of at most degree 1, we
use in each step of the differentiation
the first equation of (\ref{riemann-1}) to replace $\partial^2_xF$.
The final expressions for $\gamma'$ and $\gamma''$ are
\begin{subequations}
\begin{eqnarray}
&&\gamma'=3\rho\left(-\rho+\frac{4(\partial_xF)^2}{F^2}\right),\\
&&\gamma''=\frac{6\rho(7\rho F^2-10(\partial_xF)^2)\partial_xF}{F^3}. 
\end{eqnarray}
\end{subequations}
Putting back the expressions just found for $\gamma$, $\gamma'$ 
and $\gamma''$ in the left hand side of
(\ref{kid2-expanded}) we can check explicitly that it vanishes identically. 
Thus, the initial data set  under consideration possesses a KID with the 
required properties. 

To complete our study of the initial data set (\ref{data-set}) 
we need to solve the partial differential equations (\ref{riemann-1}), 
(\ref{riemann-2}), (\ref{hamiltonian}) or at least 
show that they have non-trivial solutions. The conditions
\begin{equation}
-\frac{\partial_yF_{} \partial_xF_{}}{F_{}}+\partial^2_{xy}F_{}=0,\
-\frac{\partial_zF_{} \partial_xF_{}}{F_{}}+\partial^2_{xz}F_{}=0,
\end{equation}
entail $F(x,y,z)=\Phi(x)G(y,z)$. Inserting this in (\ref{hamiltonian})
we obtain
\begin{equation}
\frac{5\Phi'^2}{\Phi^4}-\frac{2\Phi''}{\Phi^3}=
-(\partial_yG)^2-(\partial_zG)^2+G(\partial^2_yG+\partial^2_zG).
\end{equation}
Note that this same equation can be 
obtained if we eliminate $\rho$ in the first of (\ref{riemann-1}) and 
(\ref{riemann-2}) and insert $F(x,y,z)=\Phi(x)G(y,z)$ in the resulting 
expression.  We have reduced 
the problem to solving an ordinary differential equation
and a partial differential equation
\begin{equation}
\frac{5\Phi'^2}{\Phi^4}-\frac{2\Phi''}{\Phi^3}=k,\ 
-(\partial_yG)^2-(\partial_zG)^2+G(\partial^2_yG+\partial^2_zG)=k,
\label{pde}
\end{equation}
where $k$ is a constant. The general solution of these differential equations 
remains unknown but we can check that the standard time symmetric slice of 
Schwarzschild spacetime is among its solutions. These initial 
data are given in standard spherical coordinates by the expression
\begin{equation}
h_{ij}dx^i dx^j=\left(1+\frac{m}{2r}\right)^4(dr^2+r^2 
d\theta^2+r^2\sin^2\theta d\varphi^2),\label{h_ts}\
\end{equation}
We concentrate in a region $\mathcal U\subset\mathcal S$ 
defined by $r>m/2$. The topology of $\mathcal U$ is 
$\mathbb R^3\setminus O$ where $O$ represents an open ball. The development of
this initial data is one of the static regions of Schwarzschild spacetime.
The proof of this statement can now be obtained
 as a consequence of theorem \ref{main_result} if we show
that (\ref{h_ts}) solves (\ref{pde}). To that end we need to write 
the latter in the same coordinate system as the former.
This is achieved through the coordinate 
change 
\begin{eqnarray}
&&x=r+m\log r-\frac{m^2}{4r},\\
&&y=2\cot(\theta/2)\cos\varphi,\ z=2\cot(\theta/2)\sin\varphi.
\end{eqnarray}
The resulting differential equations are
\begin{eqnarray}
&&\left(-5 r (m+2 r)\Phi'^2+4 m \Phi 
   \Phi'+2 r (m+2 r) \Phi\Phi''\right)=\nonumber\\
&&=-k\frac{(m+2 r)^5 \Phi ^4}{16 r^3},\label{spheric-1}\\
&&\frac{1}{4} \tan^2(\theta/2) ((\partial_{\phi}G_{})^2-G_{} 
(\partial^2_{\phi}G_{})-G\sin\theta\cos\theta(\partial_{\theta}G)\nonumber\\
&&\hspace{5mm}-\sin\theta (\partial_{\theta}G_{})^2
+ G(\partial^2_{\theta}G)\sin\theta)=-k,\label{spheric-2}
\end{eqnarray}
where $\Phi=\Phi(r)$, $G=G(\theta,\phi)$ and a prime means
derivative with respect to $r$. The choice 
\begin{equation}
\Phi=\frac{1}{r\left(1+\frac{m}{2r}\right)^2},\ 
G=\frac{1}{\sin^2(\theta/2)},
\end{equation}
brings (\ref{data-set}) into (\ref{h_ts}) and 
solves (\ref{spheric-1}) and (\ref{spheric-2}) for $k=-1$.
Also this choice entails 
\begin{eqnarray}
&&\rho=\frac{64mr^3}{(m+2r)^6}\Rightarrow
\gamma=\frac{768m(m-2r)r^4}{(m+2r)^9},\\
&&Y=\frac{3(m-2r)}{m^{1/3}(m+2r)}<0,\ \mbox{if}\ r>m/2,
\end{eqnarray}
which means that (\ref{new_rho})-(\ref{new_alphaineq}) hold.

\end{document}